# 3D reconstruction of a million atoms by multiple-section local-orbital tomography

**Authors:** Liangze Mao[1,2,3†], Jizhe Cui[1,2,3†] and Rong Yu[1,2,3*]

**Affiliations:**

[1]School of Materials Science and Engineering, Tsinghua University, Beijing 100084, China.

[2]MOE Key Laboratory of Advanced Materials, Tsinghua University, Beijing 100084, China.

[3]State Key Laboratory of New Ceramics and Fine Processing, Tsinghua University, Beijing 100084, China.

*Corresponding author. Email: ryu@tsinghua.edu.cn

†These authors contributed equally to this work.

**Abstract:** There exist two groups of electron microscopy methods that are capable of providing three-dimensional (3D) structural information of an object, i.e., electron tomography and depth sectioning. Electron tomography is capable of resolving atoms in all three dimensions, but the accuracy in atomic positions is low and the object size that can be reconstructed is limited. Depth sectioning methods give high positional accuracy in the imaging plane, but the spatial resolution in the third dimension is low. In this work, electron tomography and depth sectioning are combined to form a method called multiple-section local-orbital tomography, or nLOT in short. The nLOT method provides high spatial resolution and high positional accuracy in all three dimensions. The object size that can be reconstructed is extended to a million atoms. The present method establishes a foundation for the widespread application of atomic electron tomography.

**Main Text:**

Electron tomography solves the three-dimensional (3D) structure of an object by acquiring two-dimensional (2D) images of the object at different tilting angles and utilizing reconstruction methods (*1-3*). The reconstruction quality depends on the degree to what extent the imaging process is reproduced in the reconstructions. The imaging process is accurately described by solving the Schrödinger equation for the electron-object interaction. However, this can result in a forbidding computational cost. There is therefore a need to balance the accuracy of the description of the imaging process with computational efficiency.

Typical reconstruction methods (*4-9*) usually reduce the forward imaging process to a linear projection process, also known as the "linear projection assumption" (*10*). The pixel intensity of the image is treated as a linear integral of a physical quantity of the object to be reconstructed along the projection direction. High angle annular dark field scanning transmission electron microscope (HAADF-STEM) images have been widely used in electron tomography because of their good linearity and interpretability. With the development of spherical aberration-corrected electron microscopy and advanced reconstruction methods, electron tomography has reached atomic resolution and is capable of imaging atoms in crystalline and amorphous materials (*11-18*).

In HAADF-STEM, the depth of field is approximately $2*\lambda\alpha^{-2}$ (*19-21*), where $\lambda$ is the electron wavelength and $\alpha$ the convergence semi-angle. For a typical aberration-corrected electron microscope, the depth of field is about 3-15 nm (*22*, *23*). The effect of limited depth of field in imaging is manifested in two distinct ways. On the one hand, it can be used for depth-sectioning of the object, thereby providing structural information in the electron beam direction (*22-26*). On the other hand, when the object size is larger than the depth of field, objects located outside the focal plane become rapidly blurred (*27*). The blurred object information remains in the image, which negates the applicability of the "linear projection assumption." Consequently, typical reconstruction methods produce reconstructions with significant elongation artifacts. Therefore, the application of atomic electron tomography is currently limited mainly to object sizes below 10 nm. Very recently, we proposed and realized local-orbital tomography with depth-dependent interactions (dLOT in short) to account for the evolution of the electron beam along the depth direction (*28*). It extends the object size that can be reconstructed with atomic electron tomography to 200,000 atoms and improves the positional accuracy to sub-10 pm.

Here, we propose a reconstruction method that collecting multiple images at different depths of the object at each angle, based on which tomographic reconstructions are performed to obtain 3D structures. The new method is a combination of depth-sectioning and tomography (in particular the dLOT method in this study) and is named multiple-section local-orbital tomography, or nLOT in short. nLOT describes the electron probe-object interaction as a depth-dependent convolution of the electron probe with the projection of the object within the slice at the corresponding depth, allowing the decoupling of the electron probe from the object. The number of depth-sectioning images at each tilt angle is denoted by $n$. This paper is illustrated with $n = 3$ as an example. In comparison to traditional 3D reconstruction methods, the present method demonstrates an improvement in atomic position accuracy by a factor of seven with the same electron dose, and is capable of reconstructing 3D atomic model of a million atoms with $n = 3$, or more atoms with $n > 3$ when more computing resources are available.

Before we delve into describing nLOT, a bit of clarification is necessary. nLOT is not a simple combination of depth-sectioning and tomography. Depth-sectioning typically requires fine focus step (e.g. 1 nm) in the depth direction and accurate focus values to stack the focal-series to form a 3D image of the object (*21*, *23*, *29*). In nLOT, however, much coarser focus step (e.g. 5 nm) is sufficient. Atomic electron tomography typically requires fine tilt step, e.g. 2°. In nLOT, however, much coarser tilt step (e.g. 6°) is sufficient.

## *multiple-section local-orbital tomography (nLOT)*

Conventional tomography captures only in-focus images of an object at each tilt angle to ensure optimal image acquisition (*1*, *30*). In order to acquire as much data as possible, tilt series often cover more than 40 angles between ±70°. Waiting for a period of time at each angle to ensure that the stage is stabilized greatly increases the complexity of the data acquisition and the total data acquisition time.

As shown in **Fig. 1**A, nLOT employs a data acquisition method that collects multiple images with different defocus values at each tilt angle. The defocus values of these images do not need to be distributed at equal intervals, but are roughly distributed over the entire depth range of the object. As shown in **Fig. 1**B, setting the in-focus position as the focal origin, the defocus values of nLOT for n=3 can be selected as -*L*/4, 0, and *L*/4, where *L* is the maximum size of the object.

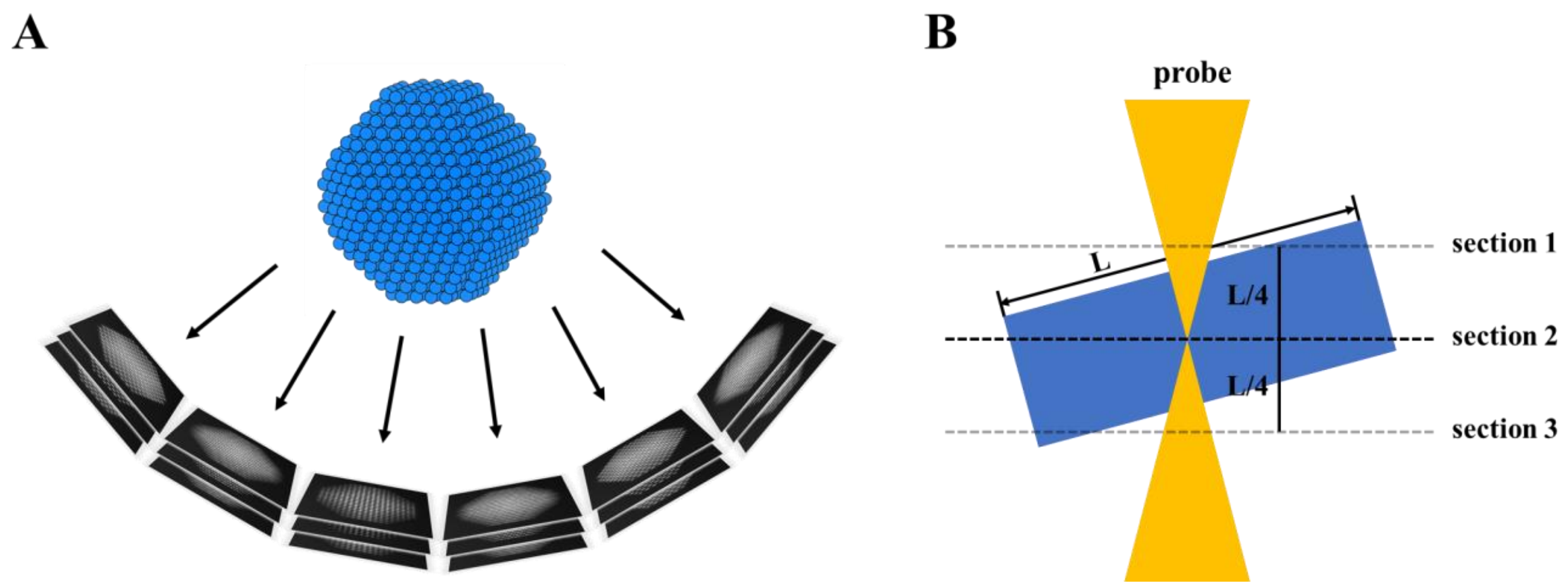

**Fig. 1. Schematic diagram of nLOT. (A)** nLOT acquires *n* images with different amounts of defocus value at each angle. n equals to 3 in the schematic. **(B)** Sections of the electron probe at different depths.

We first slice the object and the electron probe in the depth direction. By convolving the electron probe at each slice with the projection of the object within the corresponding slice and linearly accumulating it along the projection direction, the calculated image at that angle is obtained. Parameters such as atomic 3D coordinates, atomic intensity and width, sample drift, and angular deviation are directly optimized by minimizing the mean-square error between the calculated images and the recorded images. Note that the probe corresponding to each depth-section is optimized independently.

Conventional depth-sectioning typically require defocus values to be densely distributed in the depth direction of the object (*29, 31, 32*). nLOT uses a pixelated optimization approach to optimize the electron probe, which is able to decouple the electron probes from the depth-sectioning images. This allows nLOT to be very relaxed about the initial defocus value of the input dataset. With the same total dose (the same number of images in a dataset), nLOT is able to reduce the number of tilt angles by a factor of *n* compared to conventional tomography, dramatically simplifying the entire data acquisition process. Most importantly, nLOT fully utilizes the information of the entire depth range in the depth-sectioning image by considering depth-dependent interactions, a more precise forward process, and perfectly combines the advantages of depth-sectioning with those of tomography, completely breaking the limitation of depth of field on tomography.

## *Simulation tests on nLOT using amorphous particles*

To compare nLOT with single-section tomography, we generated both depth-sectioning tilt series and single-image tilt series for multiple amorphous particles using multislice simulations of HAADF-STEM images. The true coordinates of the model particles are known, facilitating the assessment of the positional accuracy of the reconstructed atomic models.

We constructed particle models containing 25,000, 50,000, 75,000, 100,000, 150,000, 200,000, 500,000, and 1,000,000 silicon atoms, named $M_1$ to $M_8$, respectively. The 3D coordinates of these silicon atoms are randomly distributed in space, and it is ensured that the nearest-neighbor distances between the atoms are maintained between 2.32 Å and 2.36 Å. As shown in **fig. S1**, the diameters of the models $M_1$ to $M_8$ are 9 nm, 12 nm, 14 nm, 15 nm, 17 nm, 19 nm, 25 nm, and 32 nm, respectively.

The tilt series $S_1$-$S_8$ are generated for single-section tomography, corresponding to the models $M_1$-$M_8$, respectively. An in-focus image was simulated at each tilt angle, which ranges from -70° to 70° with an angular interval of 2°. The parameters used for the multislice simulation include the slice thickness of 2 Å, sampling spacing of 0.1 Å for the probe and the object, zero spherical aberration, accelerating voltage of 300 kV, convergence semi-angle of 25 mrad; and detector inner and outer angles of 30 mrad and 195 mrad, respectively.

The depth-sectioning tilt series $S_1'$-$S_8'$ are generated for nLOT, corresponding to the models $M_1$-$M_8$, respectively. The same imaging parameters as $S_1$-$S_8$ are used, except the defocus values. To ensure that the nLOT datasets have the same total dose as the tilt series in the corresponding single-section tomography datasets, we adjusted the angular interval of $S_1'$ to $S_8'$ to 6°, and three images with different defocus values were simulated at each angle (defocus values of $-L/4$, 0, and $L/4$, respectively; $L$ is the particle diameter). The detailed parameters of the tilt series $S_1$ to $S_8$ and $S_1'$ to $S_8'$ are shown in Supplementary Table 1. The tilt angle distribution of the tilt series $S_1$-$S_8$ (single-section tomography) versus $S_1'$-$S_8'$ (nLOT) is shown in **fig. S2**.

For a systematic comparison between nLOT and single-section tomography, we consider three methods of describing the propagation of electrons through an object for the latter, i.e., linear projection, rigid depth-dependent interactions (rigid DDI), and adaptive depth-dependent interactions (adaptive DDI). Linear projection assumes that the electron probe is independent of depth and is a constant, which is the scheme used in most conventional tomography methods. Rigid DDI generates an electron probe based on known imaging conditions (e.g., defocus value, spherical aberration, convergence semi-angle, etc.) and keeps the electron probe unchanged during the reconstruction process. Adaptive DDI describes the depth-dependent electron probe as pixelated functions at each depth, and optimizes the probe functions during the reconstructions. Among single-section tomography methods, rigid DDI gives the most accurate results, rigid DDI the second, and linear projection the lowest (*28*).

The positional deviations between the reconstructed and true models are not uniformly distributed in space. However, the accuracy of reconstruction is often expressed in terms of the average positional deviation of all atoms, which does not reflect the inhomogeneity of the distribution of positional deviations. As shown in **Fig. 2**A, we introduce the average depth $d_i$ as a parameter to specify different regions of the object:

$$d_i = \frac{1}{M}\sum_{j=1}^{M} z_{ij} = \frac{|\boldsymbol{r_i}|}{M}\sum_{j=1}^{M}\cos(\psi_i + \theta_j)$$

where $M$ is the total number of images, $z_{ij}$ is the coordinate z of the $i$-th atom at the rotation angle $\theta_j$ corresponding to the $j$-th image. $|\boldsymbol{r_i}|$ is the length of the vector $\boldsymbol{r_i}$ that represents the distance of the $i$-th atom from the rotation axis (x), and $\psi_i$ is the angle between $\boldsymbol{r_i}$ and the $xoz$ plane.

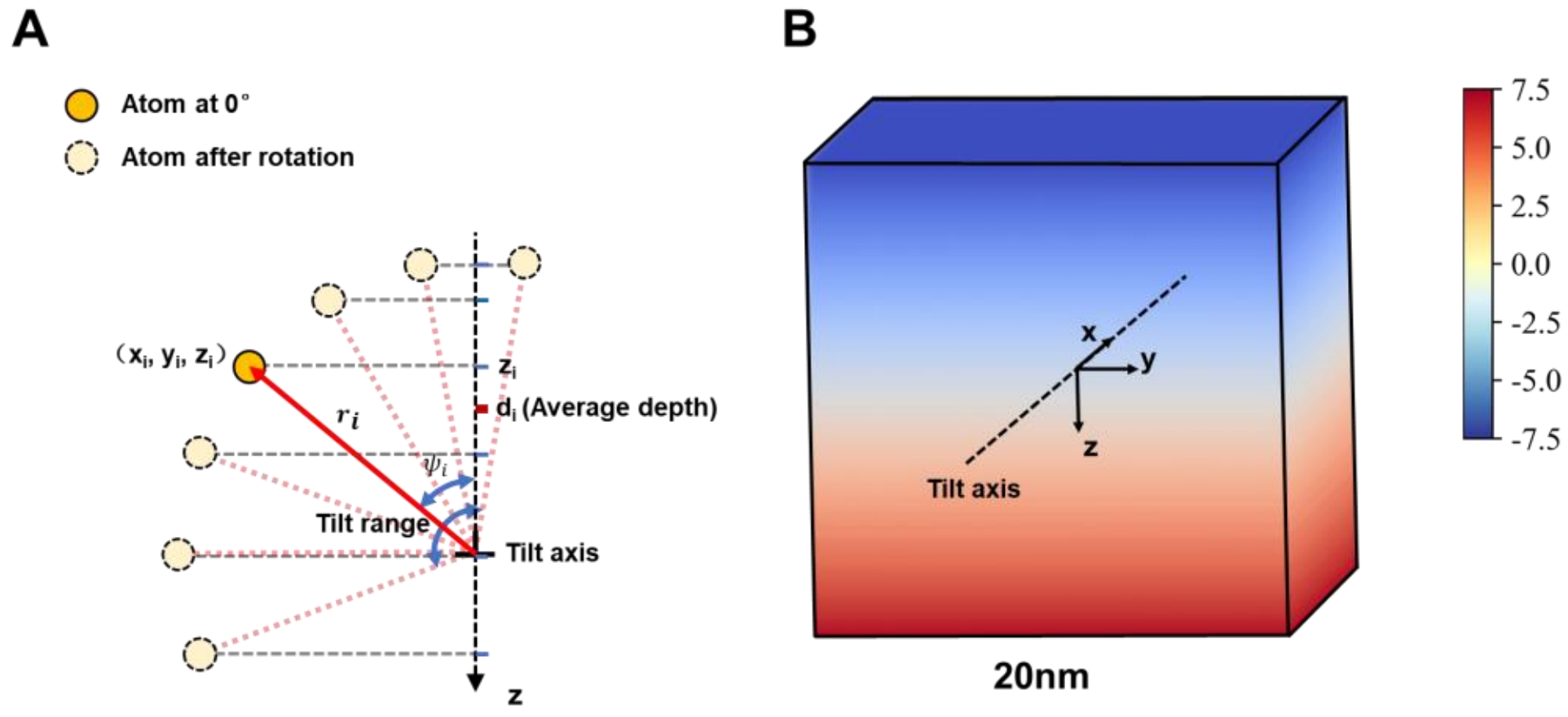

**Fig. 2. Schematic representation of the average depth. (A)** The average depth $d_i$ of an atom with coordinates $(x_i, y_i, z_i)$ at 0° corresponds to the average value of the coordinate z of the atom at each tilt angle. **(B)** Distribution of the average depth in a cubic object with a side length of 20 nm. Regions of the object with the same coordinate z have the same average depth.

When the rotation angle $\theta$ is symmetrically distributed around 0°, $\sum_{j=1}^{M} sin\theta_j = 0$. At this point, the average depth simplifies to:

$$d_i = \frac{r_i}{M}\cos\psi_i\sum_{j=1}^{M} cos\theta_j - \frac{r_i}{M}sin\psi_i\sum_{j=1}^{M} sin\theta_j = \frac{z_i}{M}\sum_{j=1}^{M} cos\theta_j$$

where $\sum_{j=1}^{M} cos\theta_j$ is a constant independent of $i$ and related only to the angular distribution of the tilt series.

It can be seen that when the rotation angle is symmetrically distributed around 0°, atoms with the same z-coordinate at 0° will also have the same average depth. The distribution of the average depth is shown in **Fig. 2**B. Since the average depth $d_i$ is proportional to the z-coordinate of the atoms at 0°, we will use the more intuitive z-coordinate of the atoms at 0° to represent their depths during tilting in the subsequent discussion.

In the following we first consider the reconstruction results of the tilt series $S_6$ using single-section tomography with adaptive DDI and $S_6$' using nLOT, respectively. The deviations of the reconstructed atomic positions from the true atomic positions were calculated in the depth direction at 4 Å intervals (the same intervals as for the probe and object sections) and plotted as box plots, as shown in **Fig. 3**A and **Fig. 3**B.

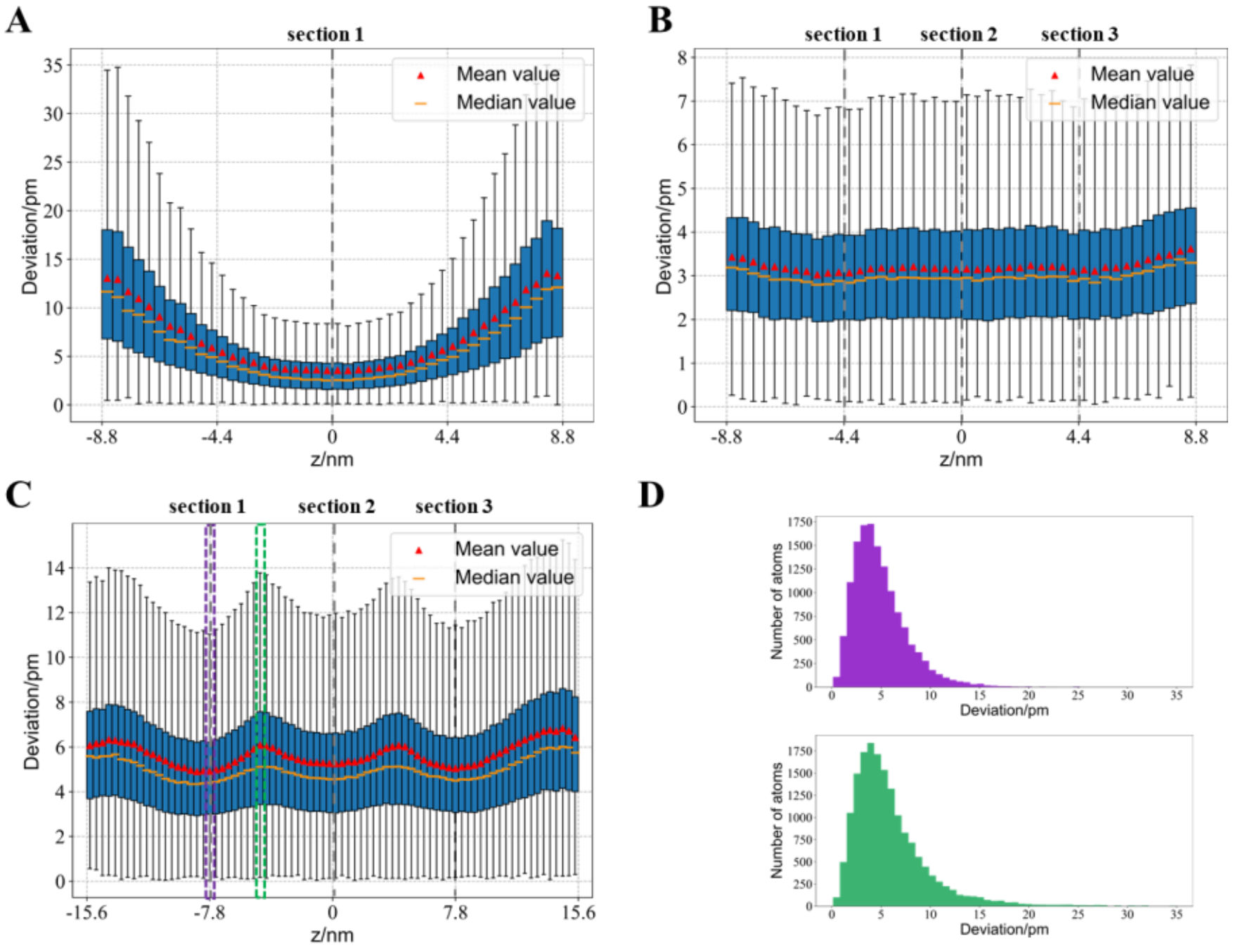

**Fig. 3. Coordinate deviation distributions of the tilt series $S_6$, $S_6$' and $S_8$'.** (**A**), (**B**) and (**C**) are box plots of coordinate deviation with depth z obtained by adaptive DDI reconstruction of tilt series $S_6$, nLOT reconstruction of tilt series $S_6$' and $S_8$', respectively. The length of the interval (width of the box plot) of the horizontal coordinate z of the box plot is all 4 Å. (**D**) Histograms of the coordinate deviations of the z coordinates in the intervals [-8.0 nm, -7.6 nm] (purple) and [-4.8 nm, -4.4 nm] (green) in **(C)**.

From **Fig. 3**A, it can be seen that the single-section tomography results in a "U" shaped positional deviation as a function of z. The minimum value in the "U" curve occurs at $z$ = 0, where the probe is focused. The spread of the probe has the smallest value at $z$ = 0, and increases with the defocus, indicating that the spreading of the probe leads to a decrease in positional accuracy. **Fig. 3**B shows the results for the tilt series $S_6$' reconstructed using nLOT with $n$ = 3. The positional deviations are small and their distribution is narrow. The distribution with depth is relatively flat and can be regarded as the stacking of three "U" shaped distributions corresponding to the three defocus values (-$L$/4, 0, and $L$/4).

**Figure 3**C shows results for the tilt series $S_8$' reconstructed using nLOT with $n$ = 3. Compared with **Fig. 3**B, the stacking of the three "U"-shaped distributions is more apparent. The local minima of the deviation distributions match the three defocus values more clearly. In addition,

the deviation distributions in each depth interval are close to the Poisson distribution, i.e., the deviations are mainly concentrated around the mean value.

The depth sections of the initial and optimized electron probe via nLOT for the tilt series $S_8$' are shown in **Fig. 4**. The initial probe is calculated based on known imaging conditions (e.g., parameters such as acceleration voltage, convergence semi-angle, etc.). Initial defocus value is set to 0 nm for all depth-sectioning images at all tilt angles. **Fig. 4**B-**4**D show the optimized probes via nLOT reconstructions corresponding to the three sections. The depths with minimum probe spread coincide with the true defocus values used for the generation of the tilt series. It can be seen that nLOT reconstructions do not depend on initial defocus values (*23*, *29*). At the time of data acquisition, it is sufficient to allow the $n$ defocus values to be roughly distributed over the thickness of the object.

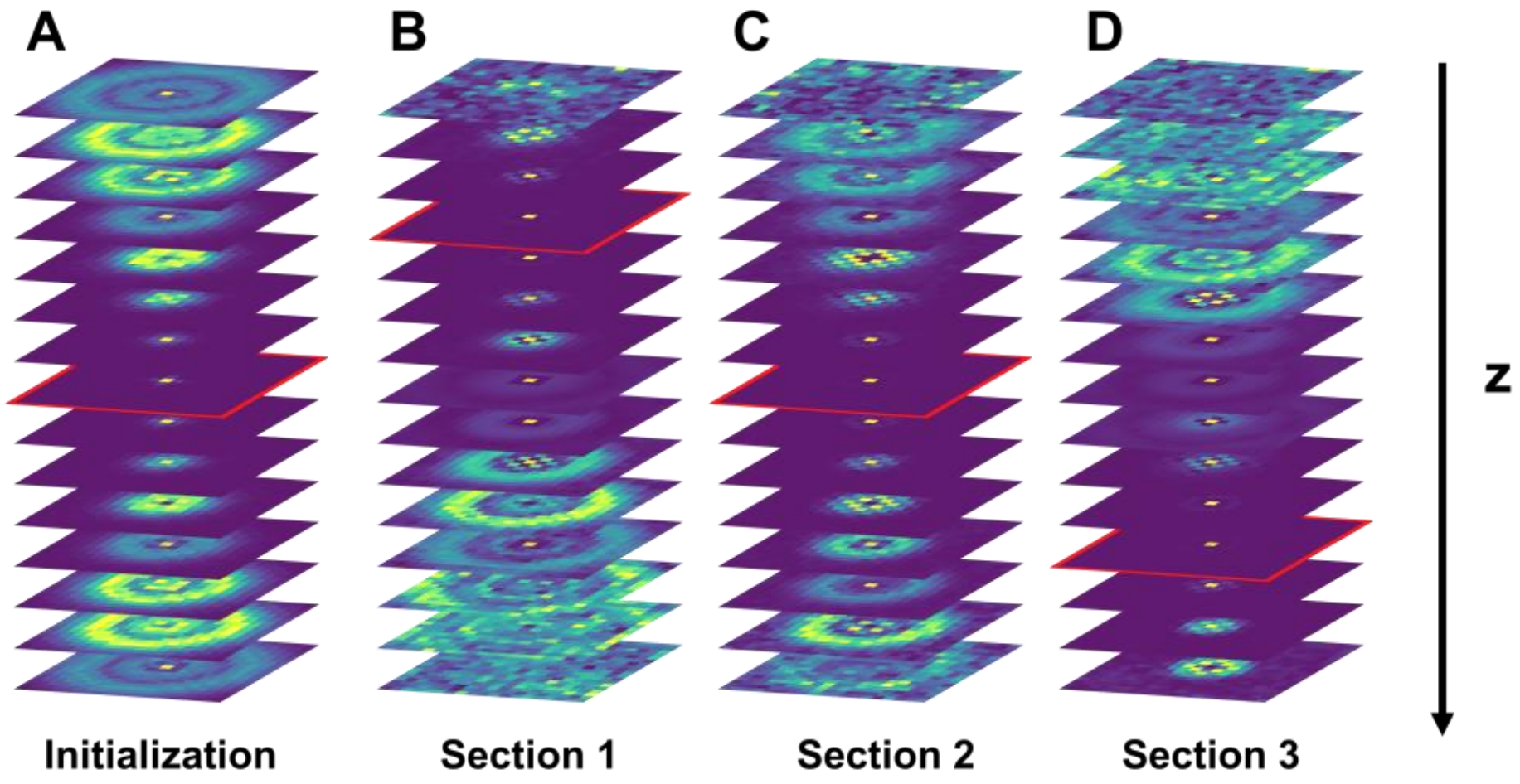

**Fig. 4. Probes of the tilt series $S_8$' reconstructed via nLOT. (A)** The depth sections of the initial probe. (**B)-(D)** The depth sections of the optimized probe corresponding to sections 1-3, respectively. The probe sections with the smallest spread are indicated by the red wireframe.

We reconstructed $S_1$'-$S_8$' using nLOT and $S_1$-$S_8$ using direct projection, rigid DDI and adaptive DDI, and the positional accuracy (RMSD) in the reconstruction results are shown in **Fig. 5**. It can be seen that the positional accuracy in the reconstructed models is nearly linear with the number of atoms for the four reconstruction methods. The slope of nLOT is significantly smaller than the slopes of the other three methods. Especially worth pointing out is the successful reconstruction of the models with 500,000 and 1000,000 atoms, which were not possible for single-section tomography methods. In particular, for the model $M_8$, which consists of 1 million atoms, nLOT achieves an accuracy of 6.6 pm. Setting 10 pm as the tolerance in positional accuracy, the number of atoms in an object that can be well reconstructed is about 75,000 for single-section tomography with the linear projection assumption, 370,000 for adaptive DDI, and over 1000,000 for nLOT with $n$ = 3.

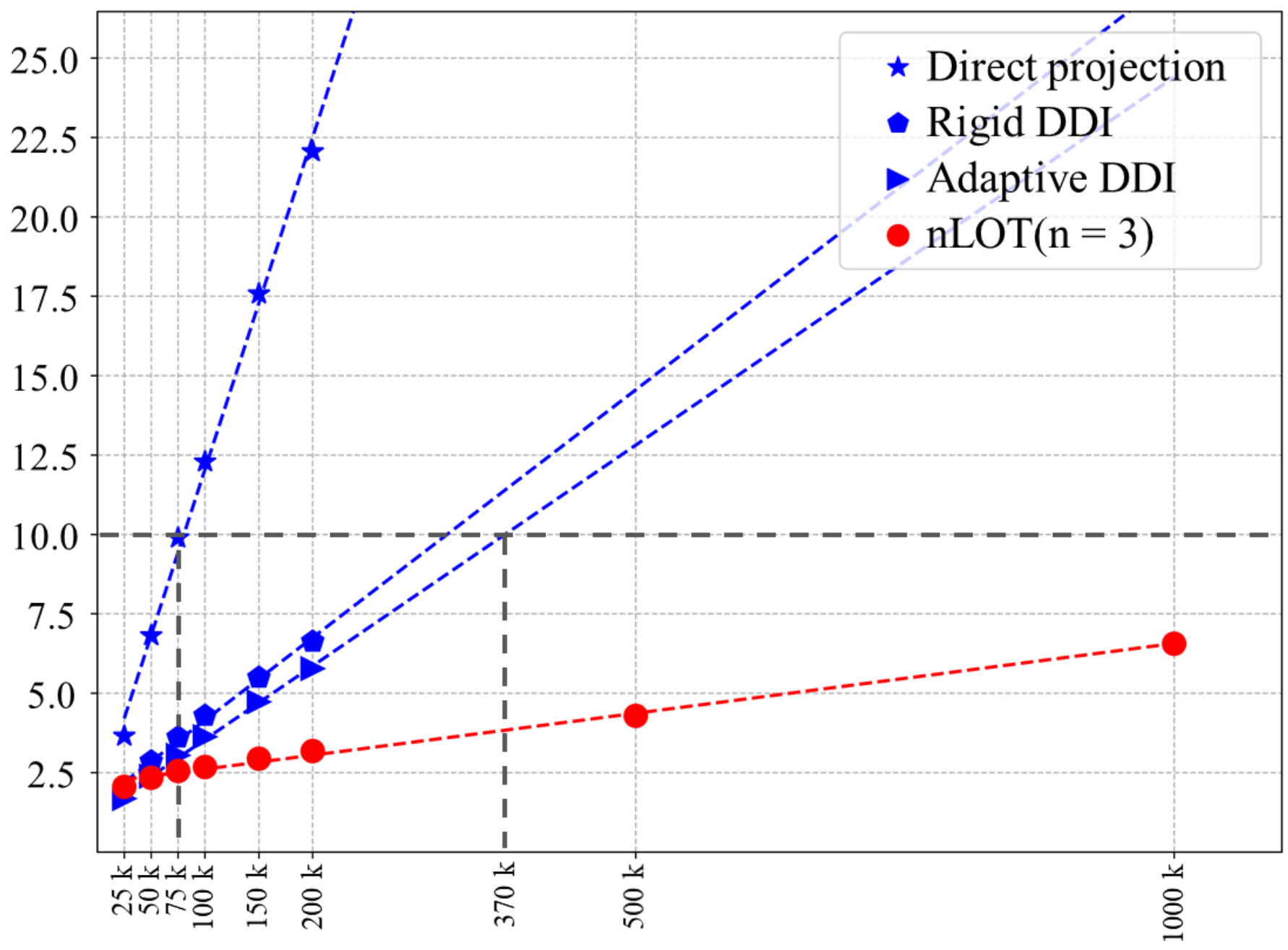

**Fig. 5. Positional accuracy of the reconstruction results for the models $M_1$ to $M_8$.** The dashed lines are linear fits to the RMSD values, with the slope being 105, 26, 23, and 4.4 pm/million atoms, for direct projection, rigid DDI, adaptive DDI, and nLOT, respectively.

Usually, image distortions and noise exist in experimental HAADF-STEM images. They vary from case to case and would lower to some extent the positional accuracy in reconstructed results. While careful experiments help to minimize the effects of image distortion, the noise is a trade-off between positional accuracy and irradiation damage.

While we demonstrate the nLOT method using models up to a million atoms with n=3, it is necessary to note that there is no theoretical limit on the object size that can be reconstructed with nLOT. For larger sizes, nLOT is expected to achieve high accuracy with more sections ($n > 3$). Of course, larger sizes require more computing resources.

## *Conclusions*

In this paper, we propose and implement multiple-section local-orbital tomography (nLOT). The method performs reconstruction by acquiring *n* sectioned images of the object at each tilt angle with different defocus values. We successfully performed atomically resolved reconstruction of amorphous model particles consisting of up to 1 million atoms. The results show that the reconstruction accuracy is approximately proportional to the number of atoms. By decoupling the probe from the image, nLOT fully utilizes the information of each depth of the object embedded in the depth-sectioning image and does not require the input of the exact defocus value, breaking the limitation of the depth of field on the size of the reconstructed object in the conventional single-section tomography and allowing the use of fewer tilt angles. The nLOT method makes possible the atomically resolved three-dimensional reconstruction of large-sized objects with high positional accuracy. It is expected to realize atomic 3D imaging of real materials, such as engineering alloys and semiconductor devices, thus greatly expanding the scope of applications for atomic electron tomography.

**Acknowledgments:** We thank H.Z. Sha and L.H. Liu for helpful discussions about the optimization of algorithms.

**Funding:**

National Natural Science Foundation of China (52388201)

National Natural Science Foundation of China (51525102)

**Author contributions:**

Conceptualization: L.M., R.Y.

Methodology: L.M, R.Y.

Simulation: L.M., J.C.

Visualization: L.M., R.Y.

Funding acquisition: R.Y.

Project administration: R.Y.

Writing – original draft: L.M.

Writing – review & editing: L.M., R.Y.

**Competing interests:** The authors declare no competing interests.

**Data and materials availability:** Correspondence and requests for materials should be addressed to Rong Yu.

**Supplementary Materials**

Supplementary Text

Figs. S1 to S2

Tables S1

# Supplementary Materials for

## 3D reconstruction of a million atoms by multiple-section local-orbital tomography

Liangze Mao[1,2,3†], Jizhe Cui[1,2,3†] and Rong Yu[1,2,3*]
Corresponding author: ryu@tsinghua.edu.cn

**The PDF file includes:**

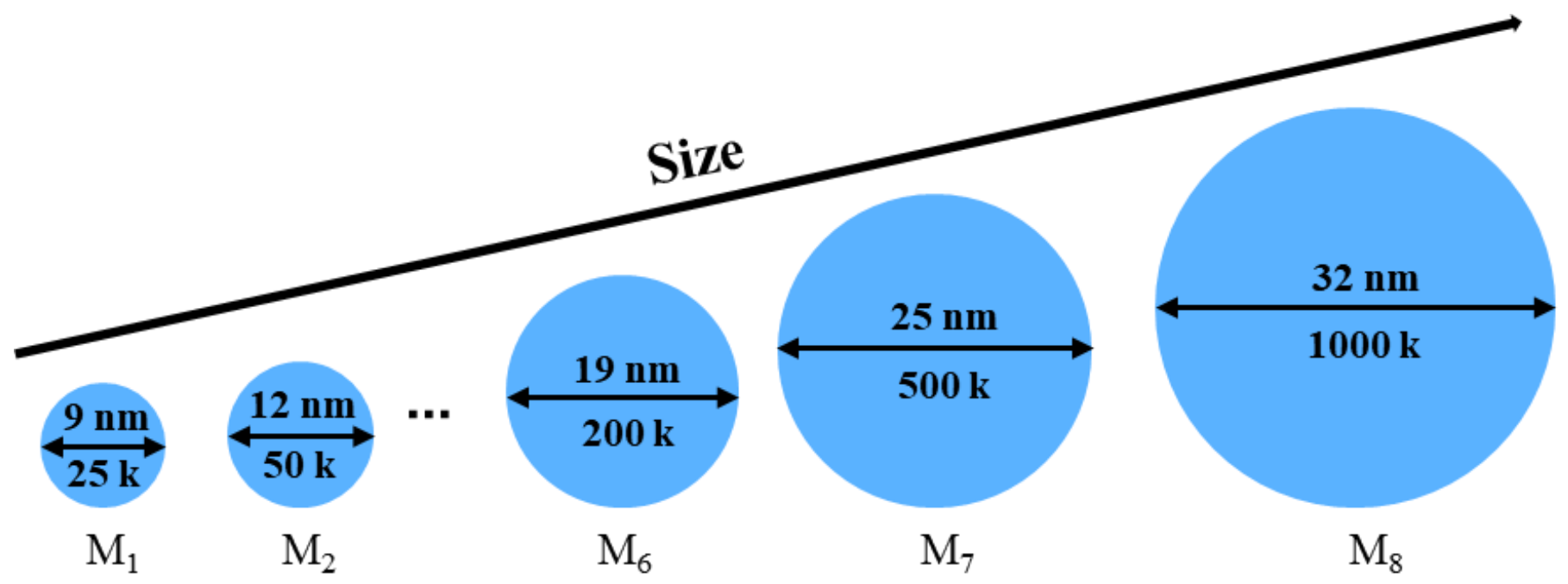

**Fig. S1.**

**Schematic diagram of model $M_1$ to $M_8$ sizes.** The diameter of the object and the number of atoms are indicated.

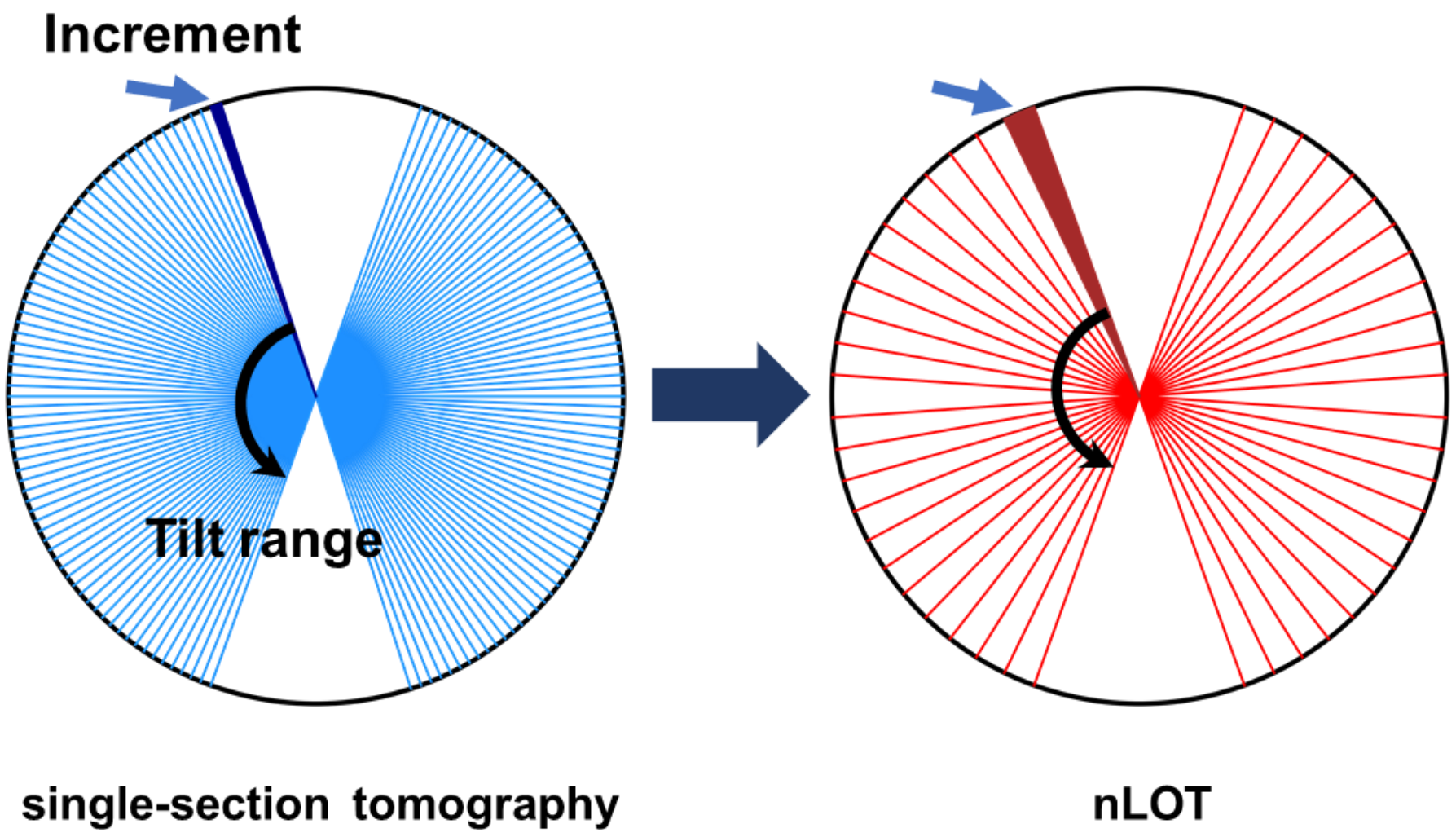

**Fig. S2.**

**Schematic diagram of the tilt angles for single-section tomography and nLOT.** The angular step is indicated.

**Table S1.**

**Parameters of $S_1$-$S_8$ and $S_1'$-$S_8'$**

| | $S_1$-$S_8$ | $S_1'$-$S_8'$ |
|---|---|---|
| Tilt range | -70.0° 70.0° | |
| # of images | 71 | |
| Pixel size | 0.34 Å | |
| Diameter | 9 nm, 12 nm, 14 nm, 15 nm, 17 nm, 19 nm, 25 nm, 32 nm | |
| # of atoms | 25 k, 50 k, 75 k, 100 k, 150 k, 200 k, 500 k, 1000 k | |
| # of defocus images/angle | 1 | 3 |